\documentclass[12pt, utf8]{article}
\usepackage[body={17cm, 23cm, centered}]{geometry}
\usepackage[english]{babel}

\usepackage[utf8]{inputenc}

\usepackage[shortcuts]{extdash}

\usepackage{amsmath,amssymb,amsthm,amscd,cite,comment,amsfonts,indentfirst,color,setspace,bbold,arydshln}
\usepackage{mathtext,marvosym,textcomp}
\usepackage{lmodern}

\usepackage[makeroom]{cancel}
\usepackage{mathtools}
\usepackage{braket}
\usepackage{cancel}
\usepackage{float}
\usepackage{bbm}

\usepackage[debug,pageanchor=false]{hyperref}
\hypersetup{colorlinks=true,linktocpage,breaklinks,
	urlcolor=blue,
	linkcolor=blue,
	citecolor=blue
}

\usepackage[vcentermath]{youngtab}

\usepackage[mathmode,centertableaux]{ytableau}
\usepackage{tikz}
\usetikzlibrary{arrows}
\usetikzlibrary{shapes.misc}
\usepackage{tikz-cd}

\usetikzlibrary{arrows,shapes,positioning, fit}
\tikzstyle{every picture}+=[remember picture]
\tikzstyle{na} = [baseline=-.5ex]
                      
\tikzset{cross/.style={cross out, draw=black, minimum size=2*(#1-\pgflinewidth), inner sep=0pt, outer sep=0pt},
	cross/.default={5pt}}
\usepackage[ normalem]{ulem}
\usepackage{tocloft}

\newcommand\scalemath[2]{\scalebox{#1}{\mbox{\ensuremath{\displaystyle #2}}}}

\numberwithin{equation}{section}

\selectfont

\def\a{\alpha} \def\b{\beta} \def\g{\gamma} \def\d{\delta}

  \def\dt{\partial}

\def\ph{\phantom}

\newcommand\bqa {\begin{eqnarray}}
\newcommand\eqa {\end{eqnarray}}

\newcommand{\bear}{\begin{array}}
\newcommand{\enar}{\end{array}}

\newcommand{\be}{\begin{equation}}
\newcommand{\ee}{\end{equation}}
\newcommand{\bea}{\begin{eqnarray}}
\newcommand{\eea}{\end{eqnarray}}

\begin{document}
\renewcommand{\contentsname}{}
\renewcommand{\refname}{\begin{center}References\end{center}}
\renewcommand{\abstractname}{\begin{center}\footnotesize{\bf Abstract}\end{center}}

\begin{titlepage}
\ph{preprint}

\vfill

\begin{center}
   \baselineskip=16pt
   {\large \bf Non-abelian uni-vector deformations in gauged supergravities and non-abelian Einstein-Maxwell theories
   }
   \vskip 2cm
     Kirill Gubarev${}^+{}^i{}^\pi$\footnote{\tt kirill.gubarev@phystech.edu },
     Konstantin Sovit$^{e}$\footnote{\tt sovit.km20@physics.msu.ru} 
       \vskip .6cm
             \begin{small}
                          {\it
                          $^i$Institute for Information Transmission Problems, 127051, Moscow, Russia\\
                          $^\pi$Moscow Institute of Physics and Technology, 
                          Laboratory of High Energy Physics, \\
                          9, Institutskii pereulok, 141702, Dolgoprudny, Russia\\
                          $^+$Institute of Theoretical and Mathematical Physics, Moscow State University, 119991, Russia\\
                            $^{e}$ Lomonosov Moscow State University, 119991, Russia
                          } \\ 
\end{small}
\end{center}

\vfill 
\begin{center} 
\textbf{Abstract}
\end{center} 
\begin{quote}
We construct non-abelian uni-vector deformations of solutions in non-abelian Einstein\--Maxwell theories and gauged supergravities, obtained as Scherk--Schwarz reductions of general relativity and double field theory, respectively. We provide examples of deformed backgrounds for both cases. We show that non-abelian deformations in Einstein--Maxwell theories can be presented as coordinate transformations in the parent theory, general relativity, extending the result of \href{https://arxiv.org/abs/2508.09637}{arXiv:2508.09637} for abelian uni-vector deformations.
\end{quote}

\vfill
\setcounter{footnote}{0}
\end{titlepage}

\tableofcontents

\setcounter{page}{2}

\newpage

\section{Introduction}

Symmetries in theoretical physics always play a crucial role, allowing one to perform difficult calculations and obtain various properties of systems. For string and M-theory, symmetry-based approaches are particularly valuable, as they help to understand the structure of these complex theories and relate them to other physically interesting ones. Among these are T-, S- and U-dualities \cite{PhysRevLett.58.1597,MONTONEN1977117,Cremmer:1997ct,Cremmer:1998px,HULL1995109}, Poisson–Lie (non-abelian) T-duality and T-plurality \cite{Klimcik:1995jn,CtiradKlimcík_2002,KLIMCIK1995455,VonUnge:2002xjf}, Nambu–Lie (non-abelian) U-duality \cite{10.1093/ptep/ptz172,Malek:2019xrf,Sakatani:2020wah,Malek:2020hpo,PhysRevD.104.046015,Musaev:2020bwm}, abelian and non-abelian fermionic T-duality \cite{Berkovits:2008ic,Bakhmatov:2011ab,NIKOLIC2017105,Astrakhantsev:2021rhj,Astrakhantsev:2022mfs,Osten:2016dvf}, Yang–Baxter deformations \cite{Bena:2003wd,Klimcik:2002zj,Klimcik:2008eq} and generalized Yang–Baxter deformations, also known as poly-vector deformations  of supergravity solutions \cite{Ashmore:2018npi,Lunin:2005jy,Bakhmatov:2019dow,Bakhmatov:2020kul,Gubarev:2020ydf,Bakhmatov:2017joy,Bakhmatov:2018apn,Bakhmatov:2018bvp,Barakin:2024rnz,Gubarev:2024tks,Gubarev:2025hvr,Gubarev:2025qox}, which allow one to construct new vacua. A special place here is occupied by the holographic, or gauge/gravity, duality \cite{Maldacena:1997re,Horowitz:2006ct}, which makes it possible to regard all the above symmetries of M- and string theories as symmetries of dual quantum field theories as well. These potentially allow one to construct new QFTs and control their properties. This means that applying deformations on the gravity side of the correspondence is dual to adding (ir)relevant or marginal operators, thereby changing the microscopic behaviour by introducing non-commutativity or by considering a phase with a non-vanishing VEV of an operator \cite{Leigh:1995ep,Argyres:1995jj,Lunin:2005jy,vanTongeren:2015uha,Imeroni:2008cr}.

Among the symmetries listed above, poly-vector deformations of supergravity solutions take a special place. They have recently attracted particular interest, as they have been found to be related to integrability \cite{Gubarev:2023jtp,Musaev:2025zww}, kappa-symmetry and generalized supergravity \cite{Bakhmatov:2022lin,Bakhmatov:2022rjn,Gubarev:2023xaq}, the sedimentation of physical objects such as particles, strings and branes \cite{Barakin:2025jwp,Barakin:2026mxz}, and $T\bar{T}$-like deformations of sigma-models \cite{Barakin:2026rcc}. One of the most recent advances concerns uni-vector deformations of Einstein–Maxwell theories with dilaton (EM) and supergravities \cite{Gubarev:2025hvr,Gubarev:2025qox,Barakin:2026mxz}, where it was shown that such deformations correspond to the sedimentation of D0-branes and can be understood as coordinate transformations in the extended space of the parent theory. However, these deformations were restricted to the abelian case and to abelian theories, and it is interesting to generalize such uni-vector deformations to the non-abelian case. This article is devoted to this generalization.

The paper is structured as follows. In Section~\ref{GTR} we consider Scherk–Schwarz reductions of general relativity and construct non-abelian uni-vector deformations of their solutions. We study two examples of such deformations for $AdS_5$ and Euclidean $AdS_4$ spaces. Moreover, we show that these non-abelian uni-vector deformations can be understood as diffeomorphisms in the extended space of the parent general relativity theory. In Section~\ref{gaugedsugra} we consider Scherk–Schwarz reductions of double field theory that yield gauged supergravities, and we construct non-abelian uni-vector deformations of their solutions. We also consider examples: deformations of the ``flat'' space and Euclidean ``flat'' space. In Section~\ref{Concl} we conclude and discuss our results, as well as opportunities for further investigation in the context of sigma-models, holographic correspondence and integrability. All calculations, performed using the computer algebra systems Wolfram Mathematica and Cadabra \cite{Peeters:2006kp,Peeters:2018dyg}, can be found in Cadabra and Mathematica files in the GitHub repository \cite{Gubarev:2025nonabelianuni}.

\section{Non-abelian uni-vector deformations in EM theories}\label{GTR}

\subsection{Scherk-Schwarz reductions of $\mathfrak{d}+\mathfrak{n}$-dimensional GR}\label{GTRSS}

Let us consider the $\mathfrak{d}+\mathfrak{n}$-dimensional Einstein-Hilbert theory on the space with $\mathfrak{d}$ external coordinates $x^{m}$ and $\mathfrak{n}$ internal coordinates $y^{\a}$: 

\begin{equation}
    \begin{aligned}
        S = & \int d^\mathfrak{d}x \, d^\mathfrak{n}y \, \sqrt{\check{G}} \left( \check{R}_{M N} \check{G}^{M N} + \Lambda \right)  = \\
        = &\int d^\mathfrak{d}x \, d^\mathfrak{n}y \, \sqrt{G} \left( -\dfrac{1}{2} \check{F}_{A B}{}^{C}\check{F}_{C D}{}^{A} S^{B D}  - \dfrac{1}{4} \check{F}_{A B}{}^{C} \check{F}_{D F}{}^{G} S_{C G} S^{A D} S^{B F} \right.\\
        & \left. \qquad \qquad \qquad \qquad \qquad \qquad \qquad + 2 \partial_{M}{\check{F}_{A} } \check{E}_{B}{}^{M} S^{A B}- \check{F}_{A} \check{F}_{B} S^{A B} + \Lambda \right) \, , 
    \end{aligned}
\end{equation}
where $ \check{F}_{A B}{}^{C} = - 2 \check{E}_{[A}{}^{M} \check{E}_{B]}{}^{N} \partial_{M}{\check{E}^{C}{}_{N}} \, ,$ $\check{F}_{A} = \check{F}_{A B}{}^{B}$, $\check{G}_{MN} = \check{E}^{A}{}_{M} \check{E}^{B}{}_{N} S_{AB}$, $\check{G} = \det \check{G}_{MN}$, $S_{AB}$ is a flat metric and $\Lambda$ is a cosmological constant. 

After performing Scherk-Schwarz reduction \cite{Scherk:1979zr} $\check{E}_{A}{}^{M}(x,y) = U_{N}{}^{M}(y) E_{A}{}^{N}(x) $, we require that the fluxes do not depend on $y$: $\check{F}_{A B}{}^{C}(x,\cancel{y}) = F_{A B}{}^{C}(x) + E_{A}{}^{\alpha}(x) E_{B}{}^{\beta}(x) E^{C}{}_{\gamma}(x) f_{\alpha \beta}{}^{\gamma} \, $. This requirement, along with the Lorentz invariance of the action, gives the following constraints on matrices $U_{M}{}^{N}(y)$
\begin{equation}
    U_{M}{}^{N}(y) \dt_{N} E_{K}{}^{A}(x) = \dt_{M} E_{K}{}^{A}(x),
\end{equation}
\begin{equation}
    - 2 U_{[\a}{}^{M}(y) U_{\b]}{}^{N}(y) \partial_{M}{U^{-1}{}_{N}{}^{\g}}(y) = f_{\a \b}{}^{\g} = \text{const}, 
\end{equation}
and all the other components of $UU\dt U^{-1} = 0$. These restrictions only leave $U_{\alpha}{}^{\beta} \neq 0$ and $U_{m}{}^{n} = \delta_{m}{}^{n} \, .$

Using the following reduction ansatz (note that here we work in the analog of the string frame for the reduced GL($\mathfrak{d}+\mathfrak{n}$) theory)
\begin{equation}
   \begin{aligned}
      G_{M N} =  
          \begin{bmatrix}
		      g_{m n} + A_{m}{}^{\alpha} A_{n}{}^{\beta} \phi_{\alpha \beta} & A_{m}{}^{\gamma}\phi_{\beta \gamma} \\
              A_{n}{}^{\gamma} \phi_{\alpha \gamma} & \phi_{\alpha \beta} 
	     \end{bmatrix} ,   \quad
         G^{M N} =  
          \begin{bmatrix}
		      g^{m n}  & - g^{m l} A_{l}{}^{\alpha} \\
            - g^{n l} A_{l}{}^{\beta}   & \phi^{\alpha \beta}  + A_{m}{}^{\alpha} A_{n}{}^{\beta} g^{m n}
	     \end{bmatrix}
   \end{aligned}
\end{equation}
we get the following action up to a total derivative (see file non-abelian\_action\_GL(n) \cite{Gubarev:2025nonabelianuni}): 
\begin{equation} \label{reduced_action}
    \begin{aligned}
      &  S_{\textbf{red}} = \int d^{\mathfrak{d}}x \sqrt{g} \sqrt{\phi} \left( R_{m n} g^{m n} - \dfrac{1}{4} F_{m n}{}^{\alpha} F_{k l}{}^{\beta} \phi_{\alpha \beta} g^{m k} g^{n l}  - \dfrac{1}{4} D_{m} \phi_{\alpha \beta} D_{n} \phi_{\alpha_1 \beta_1} \phi^{\alpha \alpha_1} \phi^{\beta \beta_1} g^{m n}  \right. \\
        & \left. \qquad \qquad + \dfrac{1}{4} \phi^{-2} g^{m n} \partial_{m} \phi  \partial_{n} \phi - \dfrac{1}{2} \phi^{\alpha \beta} f_{\alpha \gamma}{}^{\alpha_1} f_{\beta \alpha_1}{}^{\gamma} - \dfrac{1}{4} \phi_{\alpha \beta} \phi^{\gamma \alpha_1} \phi^{\alpha_2 \alpha_3} f_{\gamma \alpha_2}{}^{\alpha} f_{\alpha_1 \alpha_3}{}^{\beta}  + \Lambda  \right) \, , 
    \end{aligned}
\end{equation}
where $ F_{m n}{}^{\alpha} = \partial_{m}{A_{n}{}^{\alpha} } -  \partial_{n}{A_{m}{}^{\alpha} } -  f_{\beta \gamma}{}^{\alpha} A_{m}{}^{\beta} A_{n}{}^{\gamma}  $ and $D_{m} \phi_{\alpha \beta} =  \partial_{m}{ \phi_{\alpha \beta }} - f_{\alpha \alpha_1}{}^{\gamma} A_{m}{}^{\alpha_1} \phi_{\gamma \beta} - f_{\beta \alpha_1}{}^{\gamma} A_{m}{}^{\alpha_1} \phi_{\alpha \gamma} \, $, $\phi = |\det \phi_{\alpha \gamma}|$, $g = |\det g_{mn}|$.

The equations of motion for this theory look like: 
\begin{equation}
    \begin{aligned}
     0 = &\   R_{m n} - \dfrac{1}{2} F_{m k}{}^{\alpha} F_{n l}{}^{\beta} \phi_{\alpha \beta} g^{k l} 
        - \dfrac{1}{2} \phi^{-1}  \nabla_{m} \nabla_{n} \phi 
        + \dfrac{1}{2}  \phi^{-1} \nabla_{l} \nabla^{l} \phi \, g_{m n} 
        + \dfrac{1}{2} \phi^{-2} \partial_{m} \phi \partial_{n} \phi   \\
     &   - \dfrac{1}{4} D_{m} \phi_{\alpha \beta} D_{n} \phi_{\gamma \alpha_1} \phi^{\alpha \gamma } \phi^{\beta \alpha_1} - \dfrac{1}{4} \phi^{-2} \partial_{l} \phi \partial^{l} \phi g_{m n} 
     - \dfrac{1}{2} L \, g_{m n}   \, ,  \\
     0 = &\ \nabla^{n} F_{n m}{}^{\alpha}   + A_{k}{}^{\gamma} F_{l m}{}^{\beta} g^{k l} f_{\beta \gamma}{}^{\alpha} - F_{m l}{}^{\beta} D_{k} \phi_{\beta \gamma} \phi^{\gamma \alpha} g^{k l} - \dfrac{1}{2} F_{m n}{}^{\alpha} \phi^{-1}  \partial_{k} \phi g^{n k} + D_{m} \phi_{\beta \gamma } \phi^{\beta \alpha_1} \phi^{\alpha \alpha_2} f_{\alpha_1 \alpha_2}{}^{\gamma} \, , \\
     0 = &\  \dfrac{1}{2}  \phi^{\alpha \alpha_1} \phi^{\beta \beta_1}  \nabla^{n} D_{n} \phi_{\alpha_1 \beta_1}  - \phi^{\alpha \alpha_1} \phi^{\beta \beta_1}  g^{m n}A_{m}{}^{\beta_2} D_{n} \phi_{\alpha_1 \gamma} f_{\beta_1 \beta_2 }{}^{\gamma}
     - \dfrac{1}{2} \phi^{\alpha \beta} \phi^{-1} \nabla_{n} \nabla^{n} \phi \\
     &     - \dfrac{1}{2} g^{m n} D_{m} \phi_{\alpha_1 \beta_1} D_{n} \phi_{\alpha_2 \beta_2} \phi^{\alpha_1 \alpha_2} \phi^{\beta_1 \alpha} \phi^{\beta_2 \beta}
     + \dfrac{1}{4} g^{m n} D_{m} \phi_{\alpha_1 \beta_1} \phi^{\alpha_1 \alpha} \phi^{\beta_1 \beta} \phi^{-1}\partial_{n} \phi
     + \dfrac{1}{4} \phi^{\alpha \beta} \phi^{-2} \partial_{m} \phi \partial^{m} \phi \\
    & - \dfrac{1}{4} F_{m n}{}^{\alpha} F_{k l}{}^{\beta} g^{m k} g^{n l}
     + \dfrac{1}{2} \phi^{\alpha \gamma} \phi^{\beta \alpha_1} f_{\gamma \alpha_2}{}^{\alpha_3} f_{\alpha_1 \alpha_3}{}^{\alpha_2} 
     - \dfrac{1}{4} \phi^{\gamma \alpha_1} \phi^{\alpha_2 \alpha_3} f_{\gamma \alpha_2}{}^{\alpha} f_{\alpha_1 \alpha_3}{}^{\beta} \\
    &  + \dfrac{1}{2} \phi_{\gamma \alpha_1} \phi^{\alpha \alpha_2} \phi^{\beta \alpha_3} \phi^{\alpha_4 \alpha_5} f_{\alpha_2 \alpha_4}{}^{\gamma} f_{\alpha_3 \alpha_5}{}^{\alpha_1}
     + \dfrac{1}{2\sqrt{g}\sqrt{\phi}} L \, \phi^{\alpha \beta } \, , 
\end{aligned}
\end{equation}
where $L$ is the Lagrangian of (\ref{reduced_action}).

\subsection{Non-abelian uni-vector deformation}\label{GTRdef}

There is a way to deform a solution of this theory. If $K_{\alpha}{}^{m}$ are the Killing vectors associated with all fields and satisfy the Lie algebra with the same structure constants $f_{\alpha \beta}{}^{\gamma}$ arising from the Scherk–Schwarz reduction, such that
\begin{equation}
    K_{\alpha}{}^{n}\partial_{n} K_{\beta}{}^{m} -   K_{\beta}{}^{n}\partial_{n} K_{\alpha}{}^{m} = f_{\alpha \beta}{}^{\gamma} K_{\gamma}{}^{m} \, , 
\end{equation}
then the deformation rule 
\begin{equation} \label{deformation_G}
    \tilde{G}_{M N} = O_{M}{}^{K} O_{N}{}^{L} G_{K L} \, 
\end{equation}
with the deformation matrix 
\begin{equation}
    \begin{aligned}
  O_{M}{}^{N} = 
  \begin{bmatrix}
		    \delta_{m}{}^{n} & 0 \\
            - K_{\alpha}{}^{n} & \delta_{\alpha}{}^{\beta}
	     \end{bmatrix}  \, ,
    \end{aligned}
\end{equation}
produces another solution. Formula (\ref{deformation_G}) gives us the following deformation rules:
  \begin{equation}\label{GTRdefrules}
      \begin{aligned}
          \tilde{\phi}_{\alpha \beta} &= \phi_{\alpha \beta} - K_{\alpha}{}^{n} A_{n}{}^{\gamma} \phi_{\gamma \beta} - K_{\beta}{}^{n} A_{n}{}^{\gamma} \phi_{\alpha \gamma } + K_{\alpha}{}^{m} K_{\beta}{}^{n} g_{m n} + K_{\alpha}{}^{m} K_{\beta}{}^{n} A_{m}{}^{\alpha_1} A_{n}{}^{\beta_1} \phi_{\alpha_1 \beta_1} \, , \\
          \tilde{A}_{m}{}^{\alpha} &= \tilde{\phi}^{\alpha \beta} \left( A_{m}{}^{\gamma} \phi_{\beta \gamma} - K_{\beta}{}^{n} g_{m n} - K_{\beta}{}^{n} A_{m}{}^{\alpha_1} A_{n}{}^{\beta_1} \phi_{\alpha_1 \beta_1} \right) \, , \\
          \tilde{g}_{m n} &= g_{m n} + A_{m}{}^{\alpha} A_{n}{}^{\beta} \phi_{\alpha \beta} -   \tilde{A}_{m}{}^{\alpha} \tilde{A}_{n}{}^{\beta} \tilde{\phi}_{\alpha \beta} \, . 
      \end{aligned}
  \end{equation}

\subsection{Examples}\label{GTRex}
\subsubsection{Euclidean $Ad\mathbb{S}_4$ space}

Let us consider the case with $f_{\alpha \beta \gamma} = \epsilon_{\alpha \beta \gamma} $. We take as the initial solution: 
\begin{equation}
    \begin{aligned}
  &  g_{m n} =     \left(
\begin{array}{cccc}
 \frac{R^2}{x_4^2} & 0 & 0 & 0 \\
 0 & \frac{R^2}{x_4^2} & 0 & 0 \\
 0 & 0 & \frac{R^2}{x_4^2} & 0 \\
 0 & 0 & 0 &  \frac{R^2}{x_4^2} \\
\end{array}
\right) \, , \quad\ \phi_{\alpha \beta} = \left(
\begin{array}{ccc}
- \frac{R^2}{6} & 0 & 0 \\
 0 & -\frac{R^2}{6} & 0 \\
 0 & 0 & -\frac{R^2}{6} \\
\end{array}
\right) \, ,  \qquad A_{m}{}^{\alpha} = 0 \, , 
    \end{aligned}
\end{equation}
with $R^2 = \dfrac{15}{\Lambda} \,$. Killing vectors are $K_{\alpha}{}^{m} = \epsilon^{\alpha \b \g} \d^{m}{}_{\b} \d_{n \g} x^{n} \, , $ where the indices $\alpha , \, \beta , \, \gamma$ are lowered and raised with euclidean metric $\delta_{\alpha \beta}$ and its inverse.
The deformed solution (we use notation $\underline{x}^2 = x_1^2+x_2^2+x_3^2$): 
\begin{equation}
    \begin{aligned}
    &    \tilde{g}_{m n} = \left(
\begin{array}{cccc}
 \frac{R^2 \left(x_4^2-6 x_1^2\right)}{x_4^4-6 \left(\underline{x}^2\right) x_4^2} & -\frac{6 R^2 x_1 x_2}{x_4^4-6 \left(\underline{x}^2\right) x_4^2} & -\frac{6 R^2 x_1 x_3}{x_4^4-6 \left(\underline{x}^2\right) x_4^2} & 0 \\
 -\frac{6 R^2 x_1 x_2}{x_4^4-6 \left(\underline{x}^2\right) x_4^2} & \frac{R^2 \left(x_4^2-6 x_2^2\right)}{x_4^4-6 \left(\underline{x}^2\right) x_4^2} & -\frac{6 R^2 x_2 x_3}{x_4^4-6 \left(\underline{x}^2\right) x_4^2} & 0 \\
 -\frac{6 R^2 x_1 x_3}{x_4^4-6 \left(\underline{x}^2\right) x_4^2} & -\frac{6 R^2 x_2 x_3}{x_4^4-6 \left(\underline{x}^2\right) x_4^2} & \frac{R^2 \left(x_4^2-6 x_3^2\right)}{x_4^4-6 \left(\underline{x}^2\right) x_4^2} & 0 \\
 0 & 0 & 0 & \frac{R^2}{x_4^2} \\
\end{array}
\right) \, , \\
& \tilde{\phi}_{\alpha \beta} = \left(
\begin{array}{ccc}
 \frac{1}{6} R^2 \left(\frac{6 \left(x_2^2+x_3^2\right)}{x_4^2}-1\right) & -\frac{R^2 x_1 x_2}{x_4^2} & -\frac{R^2 x_1 x_3}{x_4^2} \\
 -\frac{R^2 x_1 x_2}{x_4^2} & \frac{1}{6} R^2 \left(\frac{6 \left(x_1^2+x_3^2\right)}{x_4^2}-1\right) & -\frac{R^2 x_2 x_3}{x_4^2} \\
 -\frac{R^2 x_1 x_3}{x_4^2} & -\frac{R^2 x_2 x_3}{x_4^2} & \frac{1}{6} R^2 \left(\frac{6 \left(x_1^2+x_2^2\right)}{x_4^2}-1\right) \\
\end{array}
\right) \, , \\
& \tilde{A}_{m}{}^{\alpha} = \left(
\begin{array}{ccc}
 0 & \frac{6 x_3}{6 \left(\underline{x}^2\right)-x_4^2} & -\frac{6 x_2}{6 \left(\underline{x}^2\right)-x_4^2} \\
 -\frac{6 x_3}{6 \left(\underline{x}^2\right)-x_4^2} & 0 & \frac{6 x_1}{6 \left(\underline{x}^2\right)-x_4^2} \\
 \frac{6 x_2}{6 \left(\underline{x}^2\right)-x_4^2} & -\frac{6 x_1}{6 \left(\underline{x}^2\right)-x_4^2} & 0 \\
 0 & 0 & 0 \\
\end{array}
\right) \, .
    \end{aligned}
\end{equation}
The transformation is nontrivial, because the Ricci scalar of the obtained background, as well as other scalar fluxes, differs from that of the initial background, that can be found in file Non-abelian\_EMD\_euclidean\_AdS4 \cite{Gubarev:2025nonabelianuni}, along with further details of the calculations.

\subsubsection{$Ad\mathbb{S}_5$ space $\to$ ``expanding brane''}

Now let us consider the 5-dimensional case with $f_{\alpha \beta \gamma} = \epsilon_{\alpha \beta \gamma} $, $R^2 = \dfrac{24}{\Lambda} \,$. The initial solution: 
\begin{equation}
    \begin{aligned}
  &  g_{m n} =     \left(
\begin{array}{ccccc}
 \frac{R^2}{x_4^2} & 0 & 0 & 0 & 0 \\
 0 & \frac{R^2}{x_4^2} & 0 & 0 & 0 \\
 0 & 0 & \frac{R^2}{x_4^2} & 0 & 0 \\
 0 & 0 & 0 & \frac{R^2}{x_4^2} & 0 \\
 0 & 0 & 0 & 0 & -\frac{R^2}{x_4^2} \\
\end{array}
\right) \, , \quad\ \phi_{\alpha \beta} = \left(
\begin{array}{ccc}
 -\frac{R^2}{8} & 0 & 0 \\
 0 & -\frac{R^2}{8} & 0 \\
 0 & 0 & -\frac{R^2}{8} \\
\end{array}
\right) \, ,  
\quad A_{m}{}^{\alpha} = 0
    \end{aligned}
\end{equation}

Killing vectors are $K_{\alpha}{}^{m} = \epsilon^{\alpha \b \g} \d^{m}{}_{\b} \d_{n \g} x^{n}$. With a notation $\underline{x}^2 = x_1^2 + x_2^2 + x_3^2$ we obtain the deformed background:
\begin{equation}
    \begin{aligned}
    &    \tilde{g}_{m n} = \left(
\begin{array}{ccccc}
 \frac{R^2 \left(x_4^2-8 x_1^2\right)}{x_4^2(x_4^2-8 \left(\underline{x}^2\right))} & -\frac{8 R^2 x_1 x_2}{x_4^2(x_4^2-8 \left(\underline{x}^2\right))} & -\frac{8 R^2 x_1 x_3}{x_4^2(x_4^2-8 \left(\underline{x}^2\right))} & 0 & 0 \\
 -\frac{8 R^2 x_1 x_2}{x_4^2(x_4^2-8 \left(\underline{x}^2\right))} & \frac{R^2 \left(x_4^2-8 x_2^2\right)}{x_4^2(x_4^2-8 \left(\underline{x}^2\right))} & -\frac{8 R^2 x_2 x_3}{x_4^2(x_4^2-8 \left(\underline{x}^2\right))} & 0 & 0 \\
 -\frac{8 R^2 x_1 x_3}{x_4^2(x_4^2-8 \left(\underline{x}^2\right))} & -\frac{8 R^2 x_2 x_3}{x_4^2(x_4^2-8 \left(\underline{x}^2\right))} & \frac{R^2 \left(x_4^2-8 x_3^2\right)}{x_4^2(x_4^2-8 \left(\underline{x}^2\right))} & 0 & 0 \\
 0 & 0 & 0 & \frac{R^2}{x_4^2} & 0 \\
 0 & 0 & 0 & 0 & -\frac{R^2}{x_4^2} \\
\end{array}
\right) \, , \\
& \tilde{\phi}_{\alpha \beta} = \left(
\begin{array}{ccc}
 \frac{1}{8} R^2 \left(\frac{8 \left(x_2^2+x_3^2\right)}{x_4^2}-1\right) & -\frac{R^2 x_1 x_2}{x_4^2} & -\frac{R^2 x_1 x_3}{x_4^2} \\
 -\frac{R^2 x_1 x_2}{x_4^2} & \frac{1}{8} R^2 \left(\frac{8 \left(x_1^2+x_3^2\right)}{x_4^2}-1\right) & -\frac{R^2 x_2 x_3}{x_4^2} \\
 -\frac{R^2 x_1 x_3}{x_4^2} & -\frac{R^2 x_2 x_3}{x_4^2} & \frac{1}{8} R^2 \left(\frac{8 \left(x_1^2+x_2^2\right)}{x_4^2}-1\right) \\
\end{array}
\right) \, , \\
& \tilde{A}_{m}{}^{\alpha} = \left(
\begin{array}{ccc}
 0 & \frac{8 x_3}{8 \left(\underline{x}^2\right)-x_4^2} & -\frac{8 x_2}{8 \left(\underline{x}^2\right)-x_4^2} \\
 -\frac{8 x_3}{8 \left(\underline{x}^2\right)-x_4^2} & 0 & \frac{8 x_1}{8 \left(\underline{x}^2\right)-x_4^2} \\
 \frac{8 x_2}{8 \left(\underline{x}^2\right)-x_4^2} & -\frac{8 x_1}{8 \left(\underline{x}^2\right)-x_4^2} & 0 \\
 0 & 0 & 0 \\
 0 & 0 & 0 \\
\end{array}
\right) \, .
    \end{aligned}
\end{equation}
Such transformation is also nontrivial, because the Ricci scalar and $ - \dfrac{1}{4} \tilde{F}^2$ scalar of the obtained background are non-trivial and have the following expressions
\begin{equation}
   \tilde{R}_{mn} \tilde{g}^{mn} = \frac{-164 x_{4}^4+480 \underline{x}^2 x_{4}^2-1408 \left(\underline{x}^2\right)^2}{\left(R x_{4}^2-8 R \underline{x}^2\right)^2} \, ,
\end{equation}
\begin{equation}
   - \dfrac{1}{4} \tilde{F}_{m n \alpha} \tilde{F}^{m n}{}_{\beta} \tilde{\phi}^{\alpha \beta} = \frac{48 x_{4}^4-96 \left(\underline{x}^2\right) x_{4}^2+256 \left(\underline{x}^2\right)^2}{\left(R x_{4}^2-8 R \left(\underline{x}^2\right)\right)^2}
\end{equation}
Details of the calculations can be found in file Non-abelian\_EMD\_AdS5 \cite{Gubarev:2025nonabelianuni}.

It is tempting to notice that the denominator $x_4^2 - 8\,\underline{x}^2$ appearing in the deformed metric components strongly resembles the worldvolume of an ``expanding brane'', with the singularity $x_4^2 = 8\,\underline{x}^2$ playing the role of a brane horizon and dimension of its world volume is 2+1=3. While a full embedding into a brane‑like picture requires further checks, this geometric similarity suggests that the non‑abelian uni‑vector deformation could be interpreted as generating an analogue of a spherical brane expanding in the $x_4$ direction. We leave a more detailed investigation of this analogy for future work.

\subsection{Deformation as diffeomorphism in extended space}\label{GTRdiffeo}

The deformation rules (\ref{GTRdefrules}) can also be written in terms of extended interval as
\begin{equation}
 \begin{aligned}
        ds^2 = \tilde{G}_{M N}(x)dx^{M} dx^{N} & = G_{m n}(x) (dx^{m} - K_{\gamma}{}^{m} dy^{\gamma} ) (dx^{n} - K_{\gamma}{}^{n} dy^{\gamma} ) \\
        &+ 2 G_{m \alpha}(x)  (dx^{m} - K_{\gamma}{}^{m} dy^{\gamma} ) dy^{\alpha} + G_{\alpha \beta}(x) \,dy^{\alpha}dy^{\beta} \, .
 \end{aligned}
 \end{equation}

However, if we attempt to interpret this transformation in terms of the coordinate transformation
 $dx'^{m} = \dfrac{\partial{x'^{m}}}{\partial x^{n}}(dx^{n} - K_{\gamma}{}^{n}dy^{\gamma}) $, 
we encounter a consistency condition,
\begin{equation}
    \dfrac{\partial^2 x'^m}{\partial y^{[\alpha} \partial y^{\beta]}} = K_{[\alpha}{}^{n} \partial_{n} K_{\beta]}{}^{k} \dfrac{\partial x'^m}{\partial x^k} = 0 \, , 
\end{equation}
which is valid for abelian Killing algebras only. Instead, the correct approach is to consider the interval
\begin{equation}
    ds'^2 = \tilde{G}_{K L} U^{-1}{}_{M}{}^{K}(y) U^{-1 }{}_{N}{}^{L}(y) dx^{M} dx^{N} \, .
\end{equation}
We take 
\begin{equation}
    \begin{aligned}
  U_{M}{}^{N} = 
  \begin{bmatrix}
		    \delta_{m}{}^{n} & 0 \\
          0 & u_{\alpha}{}^{\beta}
	     \end{bmatrix}  \, , \qquad  U^{-1 }{}_{N}{}^{M} = 
  \begin{bmatrix}
		    \delta^{m}{}_{n} & 0 \\
          0 & u^{-1 }{}_{\beta}{}^{\alpha} 
	     \end{bmatrix} \, , 
    \end{aligned}
\end{equation}
so that the structure constants  are $f_{\alpha \beta}{}^{\gamma} = -2 \,u_{[\alpha}{}^{\gamma_1} u_{\beta]}{}^{\gamma_2} \partial_{\gamma_1} u^{-1}{}_{\gamma_2}{}^{\gamma}$. Then we have 
\begin{equation} \label{interval_deformation}
 \begin{aligned}
        ds'^2 & = \tilde{G}_{K L} U^{-1}{}_{M}{}^{K}(y) U^{-1}{}_{N}{}^{L}(y) dx^{M} dx^{N}  = G_{m n}(x) (dx^{m} - K_{\gamma}{}^{m} u^{-1}{}_{\gamma_1}{}^{\gamma} dy^{\gamma_1} ) (dx^{n} - K_{\gamma}{}^{m} u^{-1}{}_{\gamma_1}{}^{\gamma} dy^{\gamma_1} ) \\
        &+ 2 G_{m \alpha}(x)  (dx^{m} - K_{\gamma}{}^{m} u^{-1}{}_{\gamma_1}{}^{\gamma} dy^{\gamma_1} ) u^{-1}{}_{\gamma_2}{}^{\alpha} dy^{\gamma_2} + G_{\alpha \beta}(x) \,u^{-1}{}_{\gamma_1}{}^{\alpha} dy^{\gamma_1} \, u^{-1}{}_{\gamma_2}{}^{\beta} dy^{\gamma_2} \, .
 \end{aligned}
 \end{equation}
Let us consider a coordinate transformation with 
\begin{equation}
    dx'^{m} = \dfrac{\partial{x'^{m}}}{\partial x^{n}}(dx^{n} - K_{\gamma}{}^{n} u^{-1}{}_{\gamma_1}{}^{ \gamma}(y) \,dy^{\gamma_1}), \qquad dy'^{\a} = dy^{\a}.
\end{equation}
One of the consistency conditions is 
\begin{equation} \label{consistency_condition}
    \begin{aligned}
          \dfrac{\partial^2 x'^m}{\partial y^{[\alpha} \partial y^{\beta]}} & = u^{-1}{}_{[\alpha}{}^{\gamma_1} u^{-1}{}_{\beta]}{}^{\gamma_2} K_{[\gamma_1}{}^{n} \partial_{n} K_{\gamma_2 ]}{}^{k} \dfrac{\partial x'^m}{\partial x^k}  - \partial_{[\beta} u^{-1}{}_{\alpha]}{}^{\gamma} K_{\gamma}{}^{n} \dfrac{\partial x'^m}{\partial x^n}  \\
          & = \dfrac{1}{2} \dfrac{\partial x'^m}{\partial x^k} u^{-1}{}_{[\alpha}{}^{\gamma_1} u^{-1}{}_{\beta]}{}^{\gamma_2} ( 2 \, K_{[\gamma_1}{}^{n} \partial_{n} K_{\gamma_2 ]}{}^{k} - f_{\gamma_1 \gamma_2}{}^{\gamma} K_{\gamma}{}^{k}   ) =0\, . 
    \end{aligned}
\end{equation}
We see that if $2 K_{[\alpha}{}^{n} \partial_{n} K_{\beta ]}{}^{m} = f_{\alpha \beta}{}^{\gamma} K_{\gamma}{}^{m} \, ,$
the consistency condition holds. The transformation we propose is 
\begin{multline} \label{coordinate_transformation}
  \scalemath{0.88}{  X'^{M} =  \sum_{i=0}^{\infty}  (-1)^{i} \int_{0}^{y} dy_{1}^{\alpha_1} u^{-1 }{}_{\alpha_1}{}^{\beta_1}(y_1) K_{\beta_1}{}^{l_1}(x) \partial_{l_1} \left(  \dots  \int_{0}^{y_{i-1}} dy_{i}^{\alpha_i} u^{-1 }{}_{\alpha_i}{}^{\beta_i}(y_{i}) K_{\beta_i}{}^{l_i}(x) \partial_{l_i} X^{M} \right)  \, } \\ 
  =  \left[ \mathcal{P} \exp \left( - \int_{0}^{y} dy'^{\alpha} u^{-1}{}_{\alpha}{}^{ \beta}(y') K_{\beta}{}^{l}(x) \frac{\partial}{\partial x^l} \right) \right] X^{M}
\end{multline}

It is important to note that only if the condition (\ref{consistency_condition}) holds (i.e. twists $u^{\alpha}{}_{\beta}(y)$  match the Killing algebra) the integrals in (\ref{coordinate_transformation}) do not depend on the path chosen, as long as $u^{-1}{}_{\alpha}{}^{ \beta}(y') K_{\beta}{}^{l}(x) \frac{\partial}{\partial x^l}$ plays role of the flat connection, and the transformation is defined correctly. It is easy to differentiate with respect to $y$, as one only needs to differentiate the first integral, to get: 
\begin{equation}
    \dfrac{\partial x'^{m}}{\partial y^{\alpha}} = - u^{-1 }{}_{\alpha}{}^{\beta_1} K_{\beta_1}{}^{l_1} \dfrac{\partial x'^{m}}{\partial x_{l_1}} \, .
\end{equation}

Our result below   shows that the coordinate transformation~(\ref{coordinate_transformation}), under the assumption that $K_{\alpha}{}^{m}$ are Killing vectors of all the fields entering the full metric $\check{G}_{MN}(x,y)$, rotates the metric indices through the matrix $U_{K}{}^{M}(y)\,O_{S}{}^{K}(x)\,U^{-1}{}_{M'}{}^{S}(y)$. This explains why the deformation requires the metric to depend on the internal coordinates through the Scherk--Schwarz ansatz,
$\check{G}_{MN}(x,y)=G_{KS}(x)\,U^{-1}{}_{M}{}^{K}(y)\,U^{-1}{}_{N}{}^{S}(y)$.

Now we prove that the transformation (\ref{coordinate_transformation}) indeed generates the deformation (\ref{interval_deformation}). 
Let us start with the undeformed interval 
\begin{equation}
  ds^2 =   G_{M N}(x') U^{-1 }{}_{K}{}^{M}(y') U^{-1 }{}_{S}{}^{N}(y') dX'^{K} dX'^{S} \, .
\end{equation}
As an example we consider only  the component $G_{m \alpha}(x') dx'^{m} \, u^{-1 }{}_{\alpha'}{}^{\alpha}(y') dy'^{\alpha'} $, since the transformation for the other components of the metric can be performed in an analogous way:
\begin{equation}
G_{m \alpha}(x') dx'^{m} \, u^{-1 }{}_{\alpha'}{}^{\alpha}(y') dy'^{\alpha'} = G_{m \alpha}(x') \dfrac{\partial{x'^{m}}}{\partial x^{n}}(dx^{n} - K_{\gamma}{}^{n}(x) u^{-1}{}_{\gamma_1}{}^{ \gamma} dy^{\gamma_1}) u^{-1 }{}_{\alpha'}{}^{\alpha}(y)  dy^{\alpha'} \, , 
\end{equation}
where we used that $dx'^{m} = \dfrac{\partial{x'^{m}}}{\partial x^{n}}(dx^{n} - K_{\gamma}{}^{n}(x) u^{-1}{}_{\gamma_1}{}^{ \gamma} dy^{\gamma_1}) \,  $ and $dy'^{\alpha} = dy^{\alpha} \, $. 
It is then sufficient to prove that $G_{n \alpha}(x') \dfrac{\partial{x'^{n}}}{\partial x^{m}} = G_{m\alpha}(x)$:
\begin{equation}
\begin{aligned}
     &    G_{n \alpha}(x') \dfrac{\partial{x'^{n}}}{\partial x^{m}} = \\
    &  = \left( G_{n \alpha}(x) 
    + \sum_{i=1}^{\infty}  (-1)^{i} \int_{0}^{y} dy_{1}^{\alpha_1} u^{-1 }{}_{\alpha_1}{}^{\beta_1}(y_1) K_{\beta_1}{}^{l_1}(x) \partial_{l_1}    \dots  \partial_{l_{i-1}}\int_{0}^{y_{i-1}} dy_{i}^{\alpha_i} u^{-1 }{}_{\alpha_i}{}^{\beta_i}(y_{i}) K_{\beta_i}{}^{l_i}(x) \partial_{l_i} G_{n \alpha}(x)    \right)  \\
 & \times   \left( \delta_{m}{}^{n} 
    + \sum_{i=1}^{\infty}  (-1)^{i} \partial_{m} \int_{0}^{y} dy_{1}^{\alpha_1} u^{-1 }{}_{\alpha_1}{}^{\beta_1}(y_1) K_{\beta_1}{}^{l_1}(x) \partial_{l_1}   \dots \partial_{l_{i-1}} \int_{0}^{y_{i-1}} dy_{i}^{\alpha_i} u^{-1 }{}_{\alpha_i}{}^{\beta_i}(y_{i}) K_{\beta_i}{}^{n}(x)  \right)  \, . 
\end{aligned}
\end{equation}
By the method of mathematical induction it is possible to show that all terms of the same order in $u^{-1}{}_{\alpha}{}^{ \beta}(y)$ cancel each other. We will ignore the factor $(-1)^{i}$, as it is the same in each order. In the first order:
\begin{equation}
  \begin{aligned}
    &    \int_{0}^{y} dy_{1}^{\alpha_1} u^{-1 }{}_{\alpha_1}{}^{\beta_1}(y_{1}) K_{\beta_1}{}^{l_1}(x) \partial_{l_1} G_{m \alpha}(x) + \int_{0}^{y} dy_{1}^{\alpha_1} u^{-1 }{}_{\alpha_1}{}^{\beta_1}(y_{1}) \partial_{m} K_{\beta_1}{}^{n}(x)   G_{n \alpha}(x) =   \\
    & \int_{0}^{y} dy_{1}^{\alpha_1} u^{-1 }{}_{\alpha_1}{}^{\beta_1}(y_{1}) \left( K_{\beta_1}{}^{l_1}(x) \partial_{l_1} G_{m \alpha}(x) + \partial_{m} K_{\beta_1}{}^{n}(x)   G_{n \alpha}(x)  \right)= 0 \, .
  \end{aligned}
\end{equation}
To continue with the arbitrary order we first introduce definitions: 
\begin{equation}
    \begin{aligned}
        R^{(i)}_{m \alpha}(x,y) & =  \int_{0}^{y} dy_{1}^{\alpha_1} u^{-1 }{}_{\alpha_1}{}^{\beta_1}(y_1) K_{\beta_1}{}^{l_1}(x) \partial_{l_1}   ( \dots ) \partial_{l_{i-1}}\int_{0}^{y_{i-1}} dy_{i}^{\alpha_i} u^{-1 }{}_{\alpha_i}{}^{\beta_i}(y_{i}) K_{\beta_i}{}^{l_i}(x) \partial_{l_i} G_{n \alpha}(x) \, , \quad i>0 \, , \\
         R^{(0)}_{m \alpha}(x) & = G_{m \alpha}(x) \, , \\
         T^{(i) n}_{m}(x,y) & = \partial_{m} \int_{0}^{y} dy_{1}^{\alpha_1} u^{-1 }{}_{\alpha_1}{}^{\beta_1}(y_1) K_{\beta_1}{}^{l_1}(x) \partial_{l_1}  ( \dots ) \partial_{l_{i-1}} \int_{0}^{y_{i-1}} dy_{i}^{\alpha_i} u^{-1 }{}_{\alpha_i}{}^{\beta_i}(y_{i}) K_{\beta_i}{}^{n}(x) \, , \quad i>0 \, , \\
          T^{(0) n}_{m} & = \delta_{m}{}^{n} \, , 
    \end{aligned}
\end{equation}
with the recurrent properties: 
\begin{equation}
    \begin{aligned}
        R^{(i+1)}_{m}(x,y) & = \int_{0}^{y} dy_{1}^{\alpha_{1}} u^{-1 }{}_{\alpha_1}{}^{\beta_1}(y_{1}) K_{\beta_1}{}^{l_{1}}(x) \partial_{l_1} R^{(i)}_{m}(x,y_1) \, , \quad i \geq 0 \, , \\
        T^{(i+1) n}_{m}(x,y) & = \int_{0}^{y} dy_{1}^{\alpha_{1}} u^{-1 }{}_{\alpha_1}{}^{\beta_1}(y_{1}) K_{\beta_1}{}^{l_{1}}(x) \partial_{l_1} T^{(i) n}_{m}(x,y_1) \\
        & \qquad \qquad + \int_{0}^{y} dy_{1}^{\alpha_{1}} u^{-1 }{}_{\alpha_1}{}^{\beta_1}(y_{1}) \partial_{m} K_{\beta_1}{}^{l_{1}}(x) T^{(i) n}_{l_1}(x,y_1) \, , \quad i \geq 0 \, . 
    \end{aligned}
\end{equation}
Let us assume now that all terms of the $i$-th order cancel each other. That means: 
\begin{equation}
   \mathcal{S}^{(i)} \equiv \sum_{s=0}^{i} R^{(s)}_{n \alpha}(x,y) T^{(i-s)n}_{m}(x,y) = 0 \, . 
\end{equation}
Now consider the $(i+1)$-th order:
\begin{equation}
    \mathcal{S}^{(i+1)} = \sum_{s=0}^{i+1} R^{(s)}_{n \alpha}(x,y) T^{(i+1-s)n}_{m}(x,y) \, .
\end{equation}
To show that it vanishes, let us show  vanishing of  its partial derivative with respect to $y^{\alpha_1}$ : 
\begin{equation}
    \begin{aligned}
        & \dfrac{\partial \mathcal{S}^{(i+1)}}{\partial y^{\alpha_1}} = \sum_{s=1}^{i+1} K_{\beta_1}{}^{l_1} u^{-1 }{}_{\alpha_1}{}^{\beta_1} \partial_{l_1} R^{(s-1)}_{n \alpha} T^{(i+1-s) n}_{m} + \sum_{s=0}^{i} R^{(s)}_{n \alpha} K_{\beta_1}{}^{l_1} u^{-1 }{}_{\alpha_1}{}^{\beta_1}  \partial_{l_1} T^{(i-s) n}_{m} \\
        & \qquad \qquad + \sum_{s=0}^{i} R^{(s)}_{n \alpha} \partial_{m} K_{\beta_1}{}^{l_1} u^{-1 }{}_{\alpha_1}{}^{\beta_1}   T^{(i-s) l_1}_{m} \\
        & = K_{\beta_1}{}^{l_1} u^{-1 }{}_{\alpha_1}{}^{\beta_1}  \partial_{l_1} \sum_{s=0}^{i} R^{(s)}_{n \alpha}  T^{(i-s) n}_{m} + \partial_{m} K_{\beta_1}{}^{l_1} u^{-1 }{}_{\alpha_1}{}^{\beta_1}  \sum_{s=0}^{i} R^{(s)}_{n \alpha}   T^{(i-s) n}_{l_1} = 0 \, . 
    \end{aligned}
\end{equation}
Considering that all terms vanish when $y=0$ and are constant, all terms with $u^{-1 }{}_{\alpha}{}^{\beta}(y)$ cancel each other  and we have 
\begin{equation}
    G_{n \alpha}(x') \dfrac{\partial{x'^{n}}}{\partial x^{m}} = G_{m\alpha}(x) \, ,
\end{equation}
that completes the proof. Finally let us note, that in the abelian case $u^{-1}{}_{\alpha}{}^{ \beta} = \delta^{\beta}_{\alpha}$ and so $u^{-1}{}_{\alpha}{}^{ \beta}(y') K_{\beta}{}^{l}(x) \frac{\partial}{\partial x^l} = K_{\alpha}{}^{l}(x) \frac{\partial}{\partial x^l}$ and $\int_{0}^{y} dy'^{\alpha} u^{-1}{}_{\alpha}{}^{ \beta}(y') K_{\beta}{}^{l}(x) \frac{\partial}{\partial x^l} = y^{\beta} K_{\beta}{}^{l}(x) \frac{\partial}{\partial x^l}$. In this case (\ref{coordinate_transformation}) reproduces the result of \cite{Gubarev:2025hvr} for abelian uni-vector deformations.

\section{Non-abelian uni-vector deformations in gauged supergravities}\label{gaugedsugra}

\subsection{Gauged supergravities from Scherk-Schwarz reductions of DFT}\label{gaugedsugraSS}

We consider Scherk-Schwarz reduction of O($\mathfrak{d}+\mathfrak{n}$,$\mathfrak{d}+\mathfrak{n}$) double field theory \cite{Grana:2012rr}
\begin{multline}
    S_{\text{DFT}} = \int d\mathbb{X} d\mathbb{Y} e^{-2d} \left( \frac{1}{4} \check{F}_{ABC} \check{F}_{A_1 B_1 C_1} S^{AA_1} \eta^{BB_1} \eta^{CC_1} - \frac{1}{12} \check{F}_{ABC} \check{F}_{A_1 B_1 C_1} S^{AA_1} S^{BB_1} S^{CC_1} \right.\\
    \left. - \frac{1}{6} \check{F}_{ABC} \check{F}_{A_1 B_1 C_1} \eta^{AA_1} \eta^{BB_1} \eta^{CC_1} + (S^{AB} - \eta^{AB}) \check{F}_A \check{F}_B \right) \, ,
\end{multline}
where $\check{F}_{A B}{}^{C} = - 2 \check{E}_{[A}{}^{M} \check{E}_{B]}{}^{N} \partial_{M}{\check{E}^{C}{}_{N}}$, $\check{\mathcal{H}}_{MN} = \check{E}^{A}{}_{M} \check{E}^{B}{}_{N} S_{AB}$, $S_{AB}$ is a flat generalized metric, $\eta^{AB}$ is an O($\mathfrak{d}+\mathfrak{n}$,$\mathfrak{d}+\mathfrak{n}$) invariant metric. The DFT coordinates are $2 \mathfrak{d}$ coordinates $\mathbb{X}^\mathcal{M} = (x^m, \tilde{x}_m)$ and $2 \mathfrak{n}$ coordinates $\mathbb{Y}^{\bar{A}} = (y^\alpha, \tilde{y}_\alpha)$. The reduction ansätze are given by $\tilde{\partial}^m = 0$ and $\check{\mathcal{H}}_{M N} = \mathcal{H}_{K L}U^{K}{}_{M}(\mathbb{Y})U^{L}{}_{N}(\mathbb{Y})$
with
\begin{equation}
  \begin{aligned}
      &   {\mathcal{H}}_{M N} =
    \left(\begin{array}{ccc}
  g_{m n}+A_m{}^{\bar{A}} M_{\bar{A} \bar{B}}  A_{n}{}^{\bar{B} }+c_{l m} g^{l s}  c_{s n}  &- g^{n l}  c_{l m} & M_{\bar{B} \bar{C}} A_m{}^{\bar{C}} + A_{l \bar{B} }  g^{l s}  c_{s m} \\
 - g^{m l}  c_{l n}    &  g^{m n} & - g^{m l}  A_{ l \bar{B}}  \\ 
M_{\bar{A} \bar{C}} A_{n}{}^{\bar{C}} + A_{ l \bar{A}}  g^{l s}  c_{s n} & -  g^{n l}  A_{ l \bar{A}} & M_{\bar{A} \bar{B}} + A_{ l \bar{A}}  g^{l s}  A_{ s \bar{B}}
    \end{array}\right) \, , \\ 
   & U^{N}{}_{M}(\mathbb{Y}) =
    \left(\begin{array}{ccc}
\delta_{m}{}^{n}  & 0 & 0 \\
0 & \delta^{m}{}_{n} & 0  \\ 
0 & 0 &  U^{\bar{B}}{}_{\bar{A}}(\mathbb{Y})
    \end{array}\right) \, , \quad   e^{-2d} = \sqrt{g} e^{-2\phi} \, ,
  \end{aligned}
\end{equation}
where $ U^{\bar{B}}{}_{\bar{A}}(\mathbb{Y})$ is an O($\mathfrak{n}$,$\mathfrak{n}$) matrix and ${\mathcal{H}}_{M N}$, $d$ depend only on external coordinates $x^m$. The resulting action, obtained from the reduced O($\mathfrak{d}+\mathfrak{n}$,$\mathfrak{d}+\mathfrak{n}$) action, reads (see file Gauged\_supergravity\_action \cite{Gubarev:2025nonabelianuni}):
\begin{equation} \label{action gauged supergravity}
\begin{aligned}
&  S=v \int d^{\mathfrak{d}} x \sqrt{g} e^{-2 \phi}\left[ R+4 \partial_n \phi \partial^n \phi-\frac{1}{12} H_{m n k} H^{m n k}-\frac{1}{4} M^{\bar{A} \bar{B}} F_{m n \bar{A}} F^{m n}{}_{\bar{B}}  + \dfrac{1}{8}D_{m}M_{\bar{A} \bar{B}} D^{m}M^{\bar{A} \bar{B}} \right. \\
  & \left.  + \dfrac{1}{4} M^{\bar{A} \bar{B}} f_{\bar{A} \bar{A}_1 \bar{B_{1}}} f_{\bar{B}}{}^{\bar{A}_1 \bar{B_{1}}}  - \dfrac{1}{12} M^{\bar{A}_1 \bar{B}_1} M^{\bar{A}_2 \bar{B}_2} M^{\bar{A}_3 \bar{B}_3} f_{\bar{A}_1 \bar{A}_2 \bar{A}_3 } f_{\bar{B}_1 \bar{B}_2 \bar{B}_3 } - \dfrac{1}{6} f_{\bar{A} \bar{B} \bar{C}}f^{\bar{A} \bar{B} \bar{C}} \right] \, ,
\end{aligned}
\end{equation}
where $\bar{A}, \bar{B},\bar{C}$ are the O($\mathfrak{n}$,$\mathfrak{n}$) indices, $M_{\bar{A} \bar{B}}$ is an O($\mathfrak{n}$,$\mathfrak{n}$) matrix, $v$ is a constant appearing from the integration over the reduced coordinates $\mathbb{Y}$ and
\begin{equation}
    \begin{aligned}
       H_{m n k} =&\ 3\left(\partial_{[m} B_{n k]}- A_{[m \bar{A}} \partial_n A_{k]}{}^{\bar{A}}\right)+ f_{\bar{A} \bar{B} \bar{C}} A_{[m}{ }^{\bar{A}} A_n^{\bar{B}} A_{k]}{}^{\bar{C}} \, , \\
    F_{m n \bar{A}} =&\ 2 \partial_{[m} A_{n] \bar{A}} - f_{\bar{A} \bar{B} \bar{C} } A_{[m}{}^{\bar{A}} A_{n]}{}^{\bar{C}} \, , \\
     D_{m}M_{\bar{A} \bar{B}} = &   \partial_{m}M_{\bar{A} \bar{B}} - f_{\bar{A} \bar{A}_1 \bar{B}_1} A_{m}{}^{\bar{A}_1} M^{\bar{B}_1}{}_{\bar{B}} - f_{\bar{B} \bar{A}_1 \bar{B}_1} A_{m}{}^{\bar{A}_1} M^{\bar{B}_1}{}_{\bar{A}} \, .
    \end{aligned}
\end{equation}
In the following section we present non-abelian uni-vector deformations of this theory and  the conditions under which they generate solutions.

\subsection{Non-abelian uni-vector deformation}\label{gaugedsugradef}

We deform the generalized metric  with the following deformation matrix:
\begin{equation} \label{deformation_matrix}
     O_{M}{}^{N} \,    =  \left(
  \begin{array}{ccc} 
  \delta^{n}{}_m & 0 & 0\\
   - \dfrac{1}{2}\gamma^m{}_{\bar{C}} \gamma^{ n \bar{C}} & \delta_{n}{}^m & \gamma^{m \bar{B}} \\
  -\gamma^{n}{}_{\bar{A}} & 0 & \delta_{\bar{A}}{}^{\bar{B}}
   \end{array}
   \right) \, , \quad \mathcal{H}'_{M N} = O_{M}{}^{R}O_{N}{}^{S}\mathcal{H}_{R S} \, ,
\end{equation}
where we assume  $\gamma^m{}_{\bar{C}}$ to be the Killing vectors of the original background, satisfying the  algebra
\begin{equation}
    [\gamma_{\bar{A}}, \gamma_{\bar{B}}] = f_{\bar{A}\bar{B} \bar{C}} \gamma^{\bar{C}} \, . 
\end{equation}
which produces the deformation rules: 
\begin{equation}  \label{deformation rules GDFT }
    \begin{aligned}
       \ \tilde{g}^{mn} =&\ g^{mn} + \Sigma^{mr} \Sigma^{nk} (g_{rk} + A_{r\bar{C}}A_k{}_{\bar{D}} M^{\bar{C}\bar{D}} +c_{lr}g^{ls}c_{sk} )  + \gamma^{m \bar{A}} \gamma^{n \bar{B}} (M_{\bar{A} \bar{B}} + A_{l \bar{A}} g^{ls}A_{s \bar{B}}) \\
        &  -  2 g^{l(n} \gamma^{m) \bar{A}} A_{l\bar{A}} - 2 g^{l(m} \Sigma^{n)s}c_{ls} + 2 \Sigma^{(m|r} \gamma^{|n)\bar{B}} (M_{\bar{A} \bar{B}} A_{r}{}^{\bar{A}} + A_{l\bar{B}} g^{ls}c_{sr}) \, ,\\
            \tilde{c}_{m n} =&\ \tilde{g}_{m k} g^{k l}c_{ln} - \tilde{g}_{m k} \Sigma^{k r} (g_{rn} + A_{r \bar{C}} A_n{}_{\bar{D}} M^{\bar{C}\bar{D}} + c_{lr}g^{ls}c_{sn}) - \tilde{g}_{m k} \gamma^{k \bar{B}} (A_n{}^{\bar{A}} M_{\bar{A} \bar{B}} +A_{l \bar{B}} g^{ls} c_{sn}) \, , \\     
            \tilde{A}_{m \bar{A}} =&\   \tilde{g}_{m n}  g^{n l}A_{l\bar{A}} +  \tilde{g}_{m n}  \Sigma^{nr} \gamma^{k}{}_{\bar{A}} (g_{rk}+A_{r \bar{C}} A_k{}_{\bar{D}} M^{\bar{C} \bar{D}} +c_{lr}g^{ls}c_{sk})  -  \tilde{g}_{m n}  \gamma^k{}_{\bar{A}}  g^{nl}c_{lk} \\ 
        &  -  \tilde{g}_{m n} \Sigma^{n r}(M_{\bar{A} \bar{B}} A_{r}{}^{ \bar{B}}  + A_{l \bar{A}}g^{ls}c_{sr}) -  \tilde{g}_{m n}  \gamma^{n\bar{B}} (M_{\bar{A} \bar{B}} + A_{l \bar{A} } g^{ls} A_{s \bar{B}}) + \tilde{g}_{m k} \gamma^{k \bar{B}} A_{n}{}^{\bar{C}} M_{\bar{B} \bar{C}} \gamma^{n}{}_{\bar{A}} \\ 
        &  + \tilde{g}_{m k} A_{n \bar{C}} \gamma^{k \bar{C}} \gamma^{s}{}_{\bar{A}} c_{l s} g^{n l}                     \, , \\
             \tilde{M}_{\bar{A} \bar{B}}  =&\   M_{\bar{A} \bar{B}}  + A_{l\bar{A}}g^{ls}A_{s\bar{B}} - \tilde{A}_{l\bar{A}}\tilde{g}^{ls}\tilde{A}_{s\bar{B}} - 2 \gamma^r{}_{(\bar{A}} M_{\bar{B}) \bar{C}}  A_{r}{}^{\bar{C}} - 2 \gamma^r{}_{(\bar{A}} A_{l \bar{B})}g^{ls}c_{sr} \\
        & \qquad  + \gamma^{r}{}_{\bar{A}}  \gamma^k{}_{\bar{B}} (g_{rk} + A_{r \bar{C}} A_{k \bar{D}} M^{\bar{C}\bar{D}} + c_{lr} g^{ls}c_{sk} )    \, ,
    \end{aligned}
\end{equation}
We also note that, in all the successful deformations presented below, the quantity $ \gamma^m{}_{\bar{C}} \gamma^{ n \bar{C}}$ vanishes. Conversely, every attempt to deform a background for which this quantity is non-vanishing failed to yield a valid solution, including the case of heterotic supergravity. We conjecture that the presence of this term may obstruct the interpretation of the deformation as a coordinate transformation in double field theory. A detailed investigation of this possibility is left for future work.

\subsection{Examples}\label{gaugedsugraex}

\subsubsection{Euclidean ``flat'' space}

Let us consider the case with $\mathfrak{d} = 4$, $\mathfrak{n} = 3$,  ${T}_{\bar{A}} = ({T}_\a,{T}^\a) = (T_1 , T_2, T_3, T^3, T^2, T^1)$, $f_{\bar{A}\bar{B}}{}^{\bar{C}}:$ $f_{\alpha \beta}{}^{\gamma} = \epsilon_{\alpha \beta \gamma}$, $f_{\alpha \beta \gamma} = \epsilon_{\alpha \beta \gamma}$. The initial solution: 
\begin{equation}
    \begin{aligned}
  &  g_{m n} = \d_{mn} \, , \qquad\ M^{\bar{A}\bar{B}} = \delta^{\bar{A}\bar{B}} \, ,  \qquad b_{mn} = 0, \qquad A_{m \bar{A}} = 0 \, ,  \qquad \phi = \frac{x_4}{2}.
    \end{aligned}
\end{equation}
We call this space ``flat'', despite the nontrivial dilaton, since the metric and scalar fields are trivial. Killing vectors are $\gamma^{\alpha m} = 0$, $\gamma_{\alpha}{}^{m} = \epsilon^{\alpha \b \g} \d^{m}{}_{\b} \d_{n \g} x^{n}$, with notation $ \underline{x}^2 = x_1^2+x_2^2+x_3^2$ we obtain the following expression for the deformed background
\begin{equation}
    \begin{aligned}
    &    \tilde{g}_{m n} = \left(
\begin{array}{cccc}
 \frac{x_1^2+1}{\underline{x}^2+1} & \frac{x_1 x_2}{\underline{x}^2+1} & \frac{x_1 x_3}{\underline{x}^2+1} & 0 \\
 \frac{x_1 x_2}{\underline{x}^2+1} & \frac{x_2^2+1}{\underline{x}^2+1} & \frac{x_2 x_3}{\underline{x}^2+1} & 0 \\
 \frac{x_1 x_3}{\underline{x}^2+1} & \frac{x_2 x_3}{\underline{x}^2+1} & \frac{x_3^2+1}{\underline{x}^2+1} & 0 \\
 0 & 0 & 0 & 1 \\
\end{array}
\right) \, , \quad \tilde{M}^{\bar{A}\bar{B}} = 
\left(
\begin{array}{cc}
 \tilde{M}^{\alpha \beta} & 0  \\
 0 & (\tilde{M}^{-1})_{\alpha \beta} \\
\end{array}
\right), \\
& \quad \tilde{M}^{\alpha \beta} = 
\left(
\begin{array}{ccc}
 \frac{x_1^2+1}{\underline{x}^2+1} & \frac{x_1 x_2}{\underline{x}^2+1} & \frac{x_1 x_3}{\underline{x}^2+1} \\
 \frac{x_1 x_2}{\underline{x}^2+1} & \frac{x_2^2+1}{\underline{x}^2+1} & \frac{x_2 x_3}{\underline{x}^2+1} \\
 \frac{x_1 x_3}{\underline{x}^2+1} & \frac{x_2 x_3}{\underline{x}^2+1} & \frac{x_3^2+1}{\underline{x}^2+1} \\
\end{array}
\right)
 \, , \quad
 \tilde{A}_{m \bar{A}} = \left(
\begin{array}{cccccc}
 0 & 0 & 0 & -\frac{x_2}{\underline{x}^2+1} & \frac{x_3}{\underline{x}^2+1} & 0 \\
 0 & 0 & 0 & \frac{x_1}{\underline{x}^2+1} & 0 & -\frac{x_3}{\underline{x}^2+1} \\
 0 & 0 & 0 & 0 & -\frac{x_1}{\underline{x}^2+1} & \frac{x_2}{\underline{x}^2+1} \\
 0 & 0 & 0 & 0 & 0 & 0 \\
\end{array}
\right)\, ,\\
& \tilde{\phi} = \frac{1}{4} \left(2 x_4+\log \left(\frac{1}{\left(\underline{x}^2+1\right)^2}\right)\right), \quad \tilde{b}_{mn} = 0 \, , 
    \end{aligned}
\end{equation}
Transformation is again nontrivial, as the Ricci scalar and the other scalar fluxes of the obtained background are non-trivial compared to the initial background. Details of the calculations can be found in file Gauged\_sugra\_geometric\_euclid \cite{Gubarev:2025nonabelianuni}.

\subsubsection{Minkowski ``flat'' space}

Now let us consider $\mathfrak{d} = 4$, $\mathfrak{n} = 3$,  ${T}_{\bar{A}} = ({T}_\a,{T}^\a) = (T_1 , T_2, T_3, T^3, T^2, T^1) $, $f_{\bar{A}\bar{B}}{}^{\bar{C}}:$ $f_{\alpha \beta}{}^{\gamma} = \epsilon_{\alpha \beta \gamma}$, $f_{\alpha \beta \gamma} = \epsilon_{\alpha \beta \gamma}$. The initial solution: 
\begin{equation}
    \begin{aligned}
  &  g_{m n} = \eta_{mn} \, , \qquad\ M^{\bar{A}\bar{B}} = -\delta^{\bar{A}\bar{B}} \, ,  \qquad b_{mn} = 0, \qquad A_{m \bar{A}} = 0 \, ,  \qquad \phi = \frac{x_4}{2}.
    \end{aligned}
\end{equation}
Again, we refer to this space as ``flat'', despite the nontrivial dilaton, since the metric and scalar fields are trivial. Killing vectors are $\gamma^{\alpha m} = 0$, $\gamma_{\alpha}{}^{m} = \epsilon^{\alpha \b \g} \d^{m}{}_{\b} \d_{n \g} x^{n}$. The deformed solution has the following expression

\begin{equation}
    \begin{aligned}
    &    \tilde{g}_{m n} = \left(
\begin{array}{cccc}
 \frac{x_1^2-1}{\underline{x}^2-1} & \frac{x_1 x_2}{\underline{x}^2-1} & \frac{x_1 x_3}{\underline{x}^2-1} & 0 \\
 \frac{x_1 x_2}{\underline{x}^2-1} & \frac{x_2^2-1}{\underline{x}^2-1} & \frac{x_2 x_3}{\underline{x}^2-1} & 0 \\
 \frac{x_1 x_3}{\underline{x}^2-1} & \frac{x_2 x_3}{\underline{x}^2-1} & \frac{x_3^2-1}{\underline{x}^2-1} & 0 \\
 0 & 0 & 0 & -1 \\
\end{array}
\right) \, , \quad \tilde{M}^{\bar{A}\bar{B}} = 
\left(
\begin{array}{cc}
 \tilde{M}^{\alpha \beta} & 0  \\
 0 & (\tilde{M}^{-1})_{\alpha \beta} \\
\end{array}
\right), \\
&\tilde{M}^{\alpha \beta} = 
\left(
\begin{array}{ccc}
 \frac{1-x_1^2}{\underline{x}^2-1} & -\frac{x_1 x_2}{\underline{x}^2-1} & -\frac{x_1 x_3}{\underline{x}^2-1} \\
 -\frac{x_1 x_2}{\underline{x}^2-1} & \frac{1-x_2^2}{\underline{x}^2-1} & -\frac{x_2 x_3}{\underline{x}^2-1} \\
 -\frac{x_1 x_3}{\underline{x}^2-1} & -\frac{x_2 x_3}{\underline{x}^2-1} & \frac{1-x_3^2}{\underline{x}^2-1} \\
\end{array}
\right)
 \, , \quad  \tilde{A}_{m \bar{A}} = \left(
\begin{array}{cccccc}
 0 & 0 & 0 & -\frac{x_2}{\underline{x}^2-1} & \frac{x_3}{\underline{x}^2-1} & 0 \\
 0 & 0 & 0 & \frac{x_1}{\underline{x}^2-1} & 0 & -\frac{x_3}{\underline{x}^2-1} \\
 0 & 0 & 0 & 0 & -\frac{x_1}{\underline{x}^2-1} & \frac{x_2}{\underline{x}^2-1} \\
 0 & 0 & 0 & 0 & 0 & 0 \\
\end{array}
\right)\, ,\\
& \tilde{\phi} = \frac{1}{4} \left(2 x_4+\log \left(\frac{1}{\left(\underline{x}^2-1\right)^2}\right)\right), \qquad \tilde{b}_{mn} = 0.
    \end{aligned}
\end{equation}
where  $ \underline{x}^2 = x_1^2+x_2^2+x_3^2$. It exhibits true physical singularity
\begin{equation}
    \tilde{R} = 2 \left(-\frac{7}{\left(\underline{x}^2-1\right)^2}+\frac{1}{\underline{x}^2-1}-1\right) \, , 
\end{equation}
\begin{equation}
  \tilde{F}_{m n \bar{A}} \tilde{F}^{m n}{}_{\bar{B}} \tilde{M}^{\bar{A} \bar{B}}  =\frac{5}{\left(\underline{x}^2-1\right)^2}+1 \, .
\end{equation}
Details of the calculations can be found in file Gauged\_sugra\_geometric\_minkowski \cite{Gubarev:2025nonabelianuni}.

\section{Conclusion and discussion}\label{Concl}

In this work we constructed non-abelian uni-vector deformations of solutions in non-abelian Einstein--Maxwell theories and gauged supergravities, obtained respectively as Scherk--Schwarz reductions of general relativity and double field theory. Explicit examples of deformed backgrounds were presented for $AdS_5$, Euclidean $AdS_4$, and Minkowski/Euclidean ``flat'' spaces, illustrating the non-trivial nature of the transformations. We further showed that the deformations in Einstein--Maxwell theories admit an interpretation as coordinate transformations in the extended space of the parent general relativity theory, provided the twist matrices $u^\alpha{}_\beta(y)$ are chosen to match the Killing algebra of the undeformed solution. This generalizes the abelian construction of~\cite{Gubarev:2025hvr} and provides a purely geometric origin for the deformation.

A natural question is whether a similar interpretation exists for the deformations of gauged supergravities introduced in Section~\ref{gaugedsugradef}. In the double field theory framework the deformation acts on the generalized metric via the matrix~(\ref{deformation_matrix}). Despite the simplicity of this transformation, its reformulation as a diffeomorphism in the full doubled space remains an open problem. Also, in all the consistent examples we have constructed, the combination $\gamma^m{}_{\bar{C}}\gamma^{n\bar{C}}$ vanishes identically; it is therefore plausible that this condition is necessary for the deformation to be well-defined and to allow a geometric interpretation in double field theory. Clarification of this point remains an open question.

The non-abelian uni-vector deformations constructed here exhibit two distinctive features. The first is their discrete nature: the Killing vectors must satisfy the algebra
\begin{equation}
    K_{\alpha}{}^{n}\partial_{n} K_{\beta}{}^{m} -   K_{\beta}{}^{n}\partial_{n} K_{\alpha}{}^{m} = f_{\alpha \beta}{}^{\gamma} K_{\gamma}{}^{m} \, ,
\end{equation}
with fixed structure constants $f_{\alpha\beta}{}^\gamma$ that characterise the reduced theory. Consequently, the Killing vectors cannot be rescaled by an arbitrary constant to generate a continuous family of solutions, in contrast to the abelian case~\cite{Gubarev:2025hvr}. This behaviour is reminiscent of discrete duality transformations such as T-, S- and U-dualities.

The second feature is the appearance of physical singularities in the deformed backgrounds, as seen in the examples above. This could possibly has a relation to a brane sedimentation \cite{Barakin:2026mxz}, but the discrete character of the non-abelian deformations makes a direct connection less clear. It remains an intriguing question whether these singular backgrounds correspond to specific (non-abelian) dualities between physical objects, to a non-abelian version of the sedimentation mechanism, or to some other phenomenon.

A promising avenue to clarify these issues is the study of the associated one-dimensional sigma models. As a first step, one can examine the particle action on the deformed background,
\begin{equation}
    S = \dfrac{1}{2} \int d\tau \left(   e^{-1} G_{M N}(x) U^{-1 M}{}_{K}(y) U^{-1 N}{}_{S}(y) \dot{X}^{K} \dot{X}^{S}  - e m^2  \right) \, .
\end{equation}
Introducing the momenta conjugate to the internal coordinates, $q_\alpha = \partial\mathcal{L}/\partial \dot{y}^\alpha$, the dynamics of the spacetime coordinates $x^m$ decouples from $y^\alpha$ and is governed by the effective equations
\begin{equation}
    \begin{aligned}
        \ddot{x}^{m} + \Gamma_{n k}{}^{m} \dot{x}^{n} \dot{x}^{k} + F_{n k}{}^{\alpha}\dot{x}^{n} g^{m k} q_{\alpha} - A_{n}{}^{\alpha} \phi^{\beta \gamma} f_{\alpha \beta}{}^{\alpha_1}g^{m n}q_{\gamma} q_{\alpha_1} + \dfrac{1}{2}\partial_{n} \phi^{\alpha \beta}g^{m n} q_{\alpha}q_{ \beta} & = 0 \, , \\
        \dot{q}_{\alpha} + A_{m}{}^{\beta} \dot{x}^{m}\phi_{\beta \gamma} f_{\alpha \alpha_1}{}^{\gamma} (q_{\beta_1 } \phi^{\beta_1 \alpha_1} - A_{n}{}^{\alpha_1}\dot{x}^{n}) + \phi_{\beta \gamma} f_{\alpha \alpha_1}{}^{\beta} (q_{\beta_1 } \phi^{\beta_1 \gamma} - A_{m}{}^{\gamma}\dot{x}^{m}) (q_{\beta_2 } \phi^{\beta_2 \alpha_1} - A_{n}{}^{\alpha_1}\dot{x}^{n}) & = 0     \, , \\
        g_{m n} \dot{x}^{m} \dot{x}^{n} + \phi^{\alpha \beta} q_{\alpha} q_{\beta} + m^2& = 0 \, .
    \end{aligned}
\end{equation}
Remarkably, the dependence on the internal coordinates $y^\alpha$ disappears completely from these equations, since the twist matrices $u^\alpha{}_\beta(y)$ only enter through the structure constants $f_{\alpha\beta}{}^\gamma$. The resulting system describes a particle moving in the deformed background with a time-dependent effective charge $q_\alpha(\tau)$. This structure strongly suggests that the integrable properties of the original sigma models are preserved by the deformation, and that the discrete transformations can be regarded as non-abelian generalizations of the dualities similar to non-abelian T-duality.

Another important open direction is the holographic interpretation of these deformed backgrounds. The deformations modify the asymptotic structure of the spacetime and introduce non-trivial profiles for the scalar and gauge fields, which on the field theory side should correspond to the addition of operators or the switching on of vacuum expectation values. A precise identification of the dual QFT remains a task for future investigation.

Finally, we briefly discuss the broader utility of uni-vector deformations as a tool for generating new solutions in Einstein--Maxwell theories. 
EM models are widely applied across high-energy physics: they serve as holographic frameworks for QCD phenomenology \cite{Rannu:2021pcq,Arefeva:2024mtl,Lilani:2025wnd,Zeng:2025tcz,Sachan:2013zza}, play a role in cosmological settings \cite{Ghezelbash:2015dka,Dunajski:2010uv}, and appear in studies of compact objects and particle phenomenology \cite{Raza:2025uda,Junior:2021svb,DeFelice:2025vef}. 
The ability to systematically construct novel configurations in these theories is therefore valuable for charting their moduli space and identifying physically viable models. 
Uni-vector deformations enrich the existing toolbox of solution-generating techniques, which already includes methods such as those developed in \cite{Cadoni:2018pav,Vigano:2022hrg,Yazadjiev:2006ew,Gubarev:2025hvr}.

\section*{Acknowledgments}

We are grateful to Edvard Musaev for his valuable comments on the text and discussions of related topics. We also thank Sergei Barakin, Daniil Kolovertnov for discussions on related topics. The work of Kirill Gubarev has been supported by Russian Science Foundation grant RSCF-24-71-10058.

\bibliography{bib.bib}
\bibliographystyle{utphys.bst}

\end{document}